%% file: egpaper_final.tex
\documentclass[10pt,twocolumn,letterpaper]{article}

\usepackage{iccv}
\usepackage{times}
\usepackage{epsfig}
\usepackage{graphicx}
\usepackage{amsmath}
\usepackage{amssymb}

\usepackage[breaklinks=true,bookmarks=false]{hyperref}
\usepackage{algorithm}
\usepackage{algorithmic}
\usepackage{tikz}
\usepackage{microtype}
\usepackage{graphicx}
\usepackage{subfigure}
\usepackage{blindtext}
\usepackage{enumitem}
\usepackage{svg}
\usepackage{booktabs} % for professional tables
\usepackage{amsmath}
\usepackage{mathtools}
\usepackage{amsthm}
\usepackage{bm}
\usepackage[capitalize,noabbrev]{cleveref}
\input{mathcommands.tex} 
\mathtoolsset{centercolon}
\newcommand{\proc}[1]{#1}
\newtheorem{assumption}{Assumption}
\usepackage{xcolor}
\usepackage{color, colortbl}
\usepackage{tabularx}
\usepackage{wrapfig}
\usepackage{multirow}
\usepackage{breqn}

\usepackage{pifont}

\iccvfinalcopy % *** Uncomment this line for the final submission

\def\iccvPaperID{32} % *** Enter the ICCV Paper ID here

\ificcvfinal\fi

\begin{document}

%%%%%%%%% TITLE
\title{Bi-Encoder Cascades for Efficient Image Search}

\author{Robert H\"onig\thanks{Equal contribution.}\\
{\tt\small rhoenig@ethz.ch}
% For a paper whose authors are all at the same institution,
% omit the following lines up until the closing ``}''.
% Additional authors and addresses can be added with ``\and'',
% just like the second author.
% To save space, use either the email address or home page, not both
\and
Jan Ackermann\footnotemark[\value{footnote}]\\
{\tt\small ackermannj@ethz.ch}
\and
Mingyuan Chi\\
{\tt\small minyuang@ethz.ch}\\
}

\maketitle
% Remove page # from the first page of camera-ready.
\ificcvfinal\thispagestyle{empty}\fi

%%%%%%%%% ABSTRACT
\begin{abstract}
% \vspace{-10pt}
Modern neural encoders offer unprecedented text-image retrieval (TIR) accuracy, but their high computational cost impedes an adoption to large-scale image searches. To lower this cost, model cascades use an expensive encoder to refine the ranking of a cheap encoder. However, existing cascading algorithms focus on cross-encoders, which jointly process text-image pairs, but do not consider cascades of  bi-encoders, which separately process texts and images. We introduce the small-world search scenario as a realistic setting where bi-encoder cascades can reduce costs. 
We then propose a  cascading algorithm that leverages the small-world search scenario to reduce lifetime image encoding costs of a TIR system. Our experiments show cost reductions by up to 6x. % Fix english
\end{abstract}

%%%%%%%%% BODY TEXT

% \vspace{-15pt}
\section{Introduction}
% \vspace{-5pt}
Search engines are the most widely used tool for information retrieval (IR) on the internet --- Google alone processes over 8.5 billion searches a day~\cite{googlesearches}. A search engine takes as input some query $q$ and returns a list of documents $\gD$ ranked by how well they match $q$. Keyword-based search ranks results by naively matching query keywords with documents. Semantic search improves keyword-based search by matching queries to documents based on their meaning.
A fruitful domain for semantic search is text-image retrieval (TIR), where documents are images and queries are texts. New semantic search engines for TIR leverage recent advances in deep learning for processing images and natural language~\cite{evertrove, Vespa, jain2017image, mishra2021deep}. Often,
these engines use an image encoder $I$ to embed each image $d\in\gD$ into a vector $\vv_d = I(d)$  and a text encoder $T$ to embed a textual query $q$
into an vector $\vv_q = \f{T}{q}$. Then, the engines rank the set of all images $\gD$
by some similarity measure $\f{s}{\vv_q,\vv_d}$.

Encoders for TIR can be divided into bi-encoders (BEs)~\cite{clip, xvlm, vse++} and
cross-encoders (CEs)~\cite{x2lvm, lxmert, uvl}. BEs process images and texts with separate encoders $I$ and $T$, whereas CEs also add cross-connections between $I$ and $T$. Hence, CEs are more powerful but must recompute all image embeddings for new queries, making them impractical for large-scale searches. Thus, we focus on BEs.

A BE TIR system incurs computational costs at \emph{build time} and at \emph{runtime}. At \emph{build time},
the system embeds all images in $\gD$. This only happens once. Afterward, at \emph{runtime}, the system
repeatedly receives queries $q$, computes their text embedding $\vv_q$, and ranks the images in $\gD$ against $\vv_q$. During runtime, the system may receive many queries $q_1, \ldots, q_n$. We consider the \emph{asymptotic runtime costs} when $n \rightarrow \infty$. We introduce the \emph{lifetime costs} as the sum of build time and asymptotic runtime costs. Lifetime costs are a natural metric to optimize but are usually not explicitly articulated. Instead, existing literature indirectly optimizes lifetime costs by reducing the cost of encoders or individual queries.

% The lifetime costs scale with the image encoding costs and the similarity ranking costs. Modern TIR systems use neural networks for the image and text encoders.

\begin{figure}
    \centering
    % \vspace*{-3em}
    \resizebox{0.45\textwidth}{!}{%
        \input{images/architecture.tikz}
    }
    % \vspace{-1em}
    \caption{Schematic of our algorithm for a 2-level cascade $\bracket{I_s, I_l}$. In this example, encoder $I_{\textrm{s}}$ computes embeddings (leftmost four squares) $\mV^{\gD}_{\textrm{s}}$ of all $|\gD| = 4$ images at build time. At runtime, the images that correspond to $m=2$ the highest-ranking embeddings (green) are processed by encoder $I_{\textrm{l}}$ that produces embeddings $\mV^{\gD_m}_{\textrm{l}}$ of higher quality. Finally, we rerank the top-2 images with $\mV^{\gD_m}_{\textrm{l}}$ to output the $k=1$ highest-ranking images. }
    % \vspace{-1em}
    \label{fig:arch_detail}
    % \vspace{-15pt}
\end{figure}

In this work, we seek to reduce the image encoding lifetime costs of TIR systems while preserving search quality.
To this end, we measure search quality as Recall@$k$, which denotes the fraction of searches that include the desired result in the top-$k$ results. Even small increases in $k$ can significantly improve the Recall@$k$~\cite{xvlm, clip}. Hence, for $m\gg k$, the top-$k$ results of a large and expensive encoder $I_{\textrm{large}}$ are likely included in the top-$m$ results of a small and cheap encoder $I_{\textrm{small}}$. This observation leads to our main idea: \emph{At build time, use the small image encoder $I_{\textrm{small}}$ to pre-compute image embeddings for all images and store them in a  cache $\mV^{\gD}_{\textrm{small}} = \cbrace{d\in\gD \mapsto \f{I_{\textrm{small}}}{d}}$. Then, at runtime, to handle a query $q$, first retrieve the top-$m$ results $\gD_m \subset \gD$ for some $m\gg k$, then compute $\mV^{\gD_m}_{\textrm{large}}$ with $I_{\textrm{large}}$ and return the top-$k$ results. Last, add $\mV^{\gD_m}_{\textrm{large}}$ to the cache for future queries.} \Cref{fig:arch_detail} illustrates this idea.

% This idea, illustrated in  \Cref{fig:arch_detail}), naturally extends to a cascade of $r$ progressively larger encoders that compute progressively smaller sets $\mV_1\ldots,\mV_r$.

How does this approach reduce image encoding lifetime costs?
A naive TIR system computes $|\gD|$ image embeddings with $I_{\textrm{large}}$ at build time. It incurs no additional image encoding costs at runtime, so it has image encoding lifetime costs of $\abs{\gD}c_{\textrm{large}}$
 where $c_{\textrm{large}}$ is the cost of executing image encoder $I_{\textrm{large}}$.
Our approach computes $|\gD|$ image embeddings with $I_{\textrm{small}}$ at build time. At runtime, we retrieve the top-$m$ images for query $q_n$ as the set $\gD_m^n$ and run $I_{\textrm{large}}$ on all images not in the cache, that is, on $\gD_m^n \setminus \bigcup_{1\leq i < n} \gD_m^i$. Hence, after $n$ queries we have run $I_{\textrm{large}}$ for $\abs{\bigcup_{1\leq i\leq n} \gD_m^i}$ times, so the asymptotic runtime costs of image encoding are $\abs{\gD_m^\infty} c_{\textrm{large}}$ with $\gD_m^\infty = \bigcup_{i\in\sN} \gD_m^i$. $\abs{\gD_m^\infty}$ is bounded by $\abs{\gD}$.
In practice, it is possible for over 90\% of all documents in $\gD$ to never be included in any search result over the lifetime of a large-scale search engine~\cite{googlesearchstats}. Formally, we have $\abs{\gD^\infty_{m}} < p\abs{\gD}$ with $p = 10\%$. We call this the \emph{$p$-small-world search scenario}. In a $10\%$-small-world search scenario, our approach evaluates $I_{\textrm{large}}$ on less than 10\% of $\gD$ at runtime, so the lifetime image encoding costs of our approach
is bounded by  $0.1\abs{\gD}c_{\textrm{large}} + \abs{\gD}c_{\textrm{small}}$.
Thus, our approach saves costs
if $0.1\abs{\gD}c_{\textrm{large}} + \abs{\gD}c_{\textrm{small}} < \abs{\gD}c_{\textrm{large}}$, that is if $c_{\textrm{small}} < 0.9 \cdot c_{\textrm{large}}$. We ignore the additional costs of reranking the top-$m$ results because they are insignificant compared to the image encoding costs of the neural encoders
used in modern TIR systems.
% More generally, for arbitrary $p$ our approach saves costs if $c_{\textrm{small}} < p c_{\textrm{large}}$.

While our approach reduces image encoding lifetime costs, it increases image encoding costs for the first few queries
when the cache for $I_{\textrm{large}}$ embeddings is still mostly empty.
In other words, our approach increases early query latency. We construct deep cascades with more than two image encoders and show that they mitigate increases in early query latency.

In this work, we make the following contributions:
\begin{itemize}[nosep,topsep=0pt]
    \item We introduce and formalize the small-world search scenario as an
    the interesting setting for optimizing TIR.
    \item We introduce a cascading algorithm for fast TIR with bi-encoders in a small-world search scenario.
    \item We show that our algorithm reduces TIR image encoding costs on standard benchmarks by up to 6x at no reduction in search quality.
    \item We investigate the benefits of deep cascades and
    demonstrate a reduction in early query latency between 1.75x and 2x.
\end{itemize}

\section{Related Work}
\vspace{-5pt}

Practitioners may employ a variety of techniques to 
reduce the lifetime costs of a TIR system.
When images are embedded with classical, fast computer vision algorithms, the ranking with similarity measure $s$ may bottleneck the system. 
A possible remedy is approximate ranking, for example by compressing the image embeddings\cite{product-quant, composite-quant}.

For neural TIR systems, the encoder cost greatly outweighs the ranking cost. Then, the most straightforward cost reduction targets model inference, for example via low-bit precision inference~\cite{lowprecision}, teacher-student distillation into a smaller model~\cite{distillation}, or pruning of unimportant weights~\cite{optbraindamage}. The prevalence of transformer-based architectures for TIR encoders suggests specializations of these techniques, like  pruning transformer heads~\cite{posttraining, sixteen} or masking transformer inputs~\cite{li2022scaling}. 
The faster models created by any of these techniques can be leveraged by model cascades, which we discuss next. 

\vspace{-10pt}
\paragraph{Model Cascades}
Model cascading is a recurrent theme in the literature on efficient machine learning (ML) systems.
FrugalML~\cite{frugalml} minimizes access costs of ML APIs by cascading two calls to a cheap API and an expensive API.
NoScope~\cite{noscope} speeds up video object detection by splitting a reference model into a sequence of two specialized models.
Model cascades have also been applied to facial key point estimation~\cite{facialest} and pedestrian detection~\cite{pedest}.

Several methods for fast TIR with CEs have been developed:  VLDeformer~\cite{vldeformer} trains a decomposable CE that can be used as a BE at inference time with minimal loss in quality.
CrispSearch~\cite{crispsearch}, LightningDot~\cite{lightningdot} and ``Retrieve Fast, Rerank Smart''~\cite{rfrs} all introduce two-level sequences of a BE whose results can be cached for approximate inference and a CE for precise inference on a subset of the BE results. This is superficially similar to our idea, but differs in key ways:
Most importantly, our cascades of BEs reduce costs compared to a single BE because they run the most expensive image encoder only on a fraction $p$ of the image corpus $\gD$. This requires the $p$-small-world search scenario. In contrast, a cascade of a BE followed by a CE reduces costs compared to a single CE because the BE image embeddings can be pre-computed, enabling a cheap initial image ranking. Then, the CE only needs to rerank the top-$k$ images. Note that the small-world search scenario is not beneficial to BE-CE cascades because the CE needs to rerun on every new image-query pair. Conversely, cascades of BEs can only reduce computational costs with a small-world search scenario because they would eventually evaluate each BE on every image. Finally, note that BE-CE cascades are limited to 2-level cascades, whereas our approach can benefit from deeper cascades.
However, our approach can complement BE-CE cascades by replacing the BE with a cascade of BEs.

\section{Models and Methods}
% \vspace{-5pt}

\paragraph{Cascaded Search}
\label{sec:cascadedsearch}
Let $\gD$ be a set of images 
that we want to query with a cascade of BEs. We now generalize our introductory 2-level cascade [$I_{\textrm{small}}$, $I_{\textrm{large}}$] to an arbitrarily deep cascade of image encoders
$\bracket{I_{\textrm{small}}, I_1, \ldots, I_r}$ that all use the same text encoder $T$ and that increase in computational cost, that is $c_{\textrm{small}} < c_{1} < \ldots < c_{r}$. We propose \Cref{alg:cascadedsearch} to query $\gD$ by ranking all images with $I_{\textrm{small}}$ and subsequently the top $m_j$ images with $I_j$. With $r=0$, \Cref{alg:cascadedsearch} reduces to a standard uncascaded BE search. 
With $r=1$, \Cref{alg:cascadedsearch} reduces to $\bracket{I_{\textrm{small}}, I_{\textrm{large}}}$.
\vspace{-10pt}
\paragraph{Lifetime costs}
Let us first restate the $p$-small-world search scenario with respect to \Cref{alg:cascadedsearch}. 
% \vspace{-5pt}
\begin{assumption}[$p$-small-world search scenario]
\label{ass:search}
Let $\gD^i_{m_1}$ be the set of the top-$m_1$ images pushed to $\mathrm{Top}$ in Line \ref{line:inittop} of \Cref{alg:cascadedsearch} when handling query $q_i$.
 % Then, \vspace{-5pt}
 $$\abs{\bigcup_{i\in\sN}\gD^i_{m_1}} < p\abs{\gD}.$$
\end{assumption}

Under \Cref{ass:search}, the \mbox{$r\!+\!1$}-level cascade 
$[I_{\textrm{small}}, I_1, \\\ldots, I_r]$ has image encoding lifetime costs bounded by $C_{\textrm{$r\!\!+\!\!1$-level}} := \abs{\gD}c_{\textrm{small}} + p\abs{\gD}\sum_{j=1}^rc_j.$
The standard BE search uses only the largest image encoder $I_r$, so it has costs
$C_{\textrm{$1$-level}} = \abs{\gD}c_{\textrm{r}}.$
% Finally, the 2-level cascade $\bracket{I_{\textrm{small}}, I_r}$ has costs bounded by $C_{\textrm{$2$-level}} = \abs{\gD}c_{\textrm{small}} + p\abs{\gD}c_r.$
Hence, the \mbox{$r\!+\!1$}-level cascade reduces costs
by the factor 
\vspace{-5pt}
$$F_{\textrm{life}} := {C_{\textrm{$1$-level}}}/{C_{\textrm{$r\!\!+\!\!1$-level}}} = c_r / \brace{c_{\textrm{small}} + p\sum_{j=1}^rc_j}.$$  It is clear that a $2$-level cascade always achieves a greater cost reduction than a cascade with more than two levels because the denominator of $F_{\textrm{life}}$ grows with $r$.

Uncascaded models encode all images at build time, so they incur no additional image encoding costs at runtime. In contrast, cascades incur image encoding costs at runtime on cache misses. Cache misses occur most frequently for early queries. This means that all cascades experience additional latency for early queries, even though they have lower lifetime costs. We now show that deep cascades may reduce early query latency compared to $2$-level cascades. 

\begin{figure}
\vspace{-10pt}
  % \begin{minipage}{0.45\textwidth}
    % \vspace*{-2.7em}
    \begin{algorithm}[H]
     \caption{Cascaded Search. Here, $\f{\mathrm{Rank}}{\gI, \mV, \vv_q}$ sorts the images in $\gI$ by the similarity $\f{s}{\vv_d,\vv_q}$ of their image encodings $\vv_d \in \mV$ with text encoding $\vv_q$.}
     \label{alg:cascadedsearch}
  \begin{algorithmic}[1]
    \STATE {\bfseries Input:} $\bracket{I_{\textrm{small}}, I_1,\ldots,I_r}$, $m_1 > \ldots > m_r \in \sN$, $\gD$, $k$
    \STATE {\bfseries Build time:} \algorithmicfor\,\,$d \in \gD$
\algorithmicdo\,\,\,$\mV^{\gD}_{\textrm{small}}\!\bracket{d} \longleftarrow \f{I_{\textrm{small}}}{d}$
      \vspace{0.2em}
    \STATE \textbf{Function} \proc{Query}($\mathrm{text}$)
      \STATE $\mathrm{Top} \longleftarrow \f{\mathrm{Rank}}{\gD,\mV^{\gD}_{\textrm{small}},\f{T}{\mathrm{text}}}$
    \label{line:inittop}
      \FOR {$j=1$ {\bfseries to} $r$}
      \STATE \algorithmicfor\,\,$d\in\mathrm{Top}\!\bracket{1\ldots m_j}$\,\,\algorithmicdo\,\, $\mV_j\!\bracket{d} \xleftarrow{\text{if empty}} \f{I_j}{d}$ 
      \label{line:cacheaccess}
      % \vspace{-1.3em}
      \STATE $\mathrm{Top} \longleftarrow \f{\mathrm{Rank}}{ \mathrm{Top}\!\bracket{1\ldots m_j},\mV_j,\f{T}{\mathrm{text}}}$
      \ENDFOR
      \STATE {\bfseries return} $\mathrm{Top}\!\bracket{1\ldots k}$
    \STATE \textbf{EndFunction}
    \STATE {\bfseries Runtime:} 
    \FOR {$i=1$ {\bfseries to} $\infty$}
    \STATE $q_i \longleftarrow $ \proc{GetUserQuery}()
    \STATE {\bfseries print} \proc{Query}($q_i$)
    \label{line:processquery}
    \ENDFOR
  \end{algorithmic}
  \end{algorithm}
% \end{minipage}
\vspace{-20pt}
\end{figure}

\vspace{-10pt}
\paragraph{Early query latency} Consider the $2$-level cascade $\bracket{I_{\textrm{small}}, I_{\textrm{large}}}$. 
\Cref{alg:cascadedsearch} repeatedly processes user queries in Line \ref{line:processquery}.  For early queries (low $i$), we can assume that the image embedding cache $\mV_{\textrm{large}}$ is empty when accessed in Line \ref{line:cacheaccess}. Then, Query($q_i$) needs to compute all $m_{\textrm{large}}$ required image embeddings, so an early query incurs image encoding cost $c_{\textrm{large}}m_{\textrm{large}}$. We refer to this cost as the \emph{early query latency}.
We can create a deep cascade
by inserting additional image encoders $I_1, \ldots, I_{r-1}$ between $I_{\textrm{small}}$ and $I_{\textrm{large}}$.
Concretely, consider the deep cascade $\bracket{I_{\textrm{small}}, I_1, \ldots, I_{r}=I_{\textrm{large}}}$. We choose $m_r < \ldots < m_2 < m_1 = m_{\textrm{large}}$.
Hence, we use the largest encoder $I_r = I_{\textrm{large}}$ only on $m_r$ images, less than the $m_{\textrm{large}}$ uses in the $2$-level encoder. This allows the deep cascade to lower early query latency. At the same time,  $m_1 = m_{\textrm{large}}$ helps ensure that search quality does not degrade.

By how much is early query latency reduced?
Similarly to the $2$-level cascade, we can assume that for early queries, the image embedding caches $\cbrace{\mV_j}_{j=1}^r$ are empty when accessed in Line \ref{line:cacheaccess}, so Query($q_i$) incurs image encoding costs $\sum_{j=1}^rc_jm_j$. Hence, the deep cascade
lowers early query latency by a factor of 
\vspace{-5pt}
\begin{equation}
\vspace{-5pt}
\label{eq:latencyreduction}
F_{\textrm{latency}} := m_{\textrm{large}}c_{\textrm{large}}/\sum_{j=1}^rc_jm_j
\end{equation}

We reduce early query latency ($F_{\textrm{latency}} > 1$) if the cost of the newly inserted encoders, $\sum_{j=1}^{r-1}c_jm_j$, is lower than the savings from fewer uses of the large encoder, $c_{\textrm{large}}(m_{\textrm{large}} - m_r)$.

\section{Experiments}
% \vspace{-5pt}

\paragraph{Experimental Setup}
% \setlength{\parskip}{0pt}
% \paragraph{Models} We use CLIP with the ViT-B/16\footnote{ViT-B/$n$ tiles an image into $n\times n$ patches.} image encoder as our baseline 1-level cascade $\bracket{I_{\textrm{large}}}$. For retraining, we set $\f{M}{I_{\textrm{large}}}$ to
% the four times faster ViT-B/32 image encoder.
Given a dataset of image-caption pairs, we measure the Recall@$k$ (R@$k$) metric as the fraction of captions whose corresponding image is among the top-$k$ search results. In line with the IR literature, we report the Recall@$k$ for $k\in\cbrace{1,5,10}$ on the Flickr30k~\cite{flickr30k} test set with 1k samples and the MSCOCO~\cite{mscoco} validation set with 5k. In addition, we report the lifetime cost reduction for 2-level cascades and the early query latency reduction for 3-level cascades. Our lifetime cost computation assumes a $p$-small-world search scenario with $p=0.1$ and $m_1 = 50$. We measure image encoding costs as the number of Multiply-Accumulate (MAC) operations reported by PyTorch-OpCounter~\cite{thop}.

We demonstrate our cascades on the popular and competitive OpenCLIP~\cite{openclip} and BLIP~\cite{blip} text-image BEs. 
Both models match images to texts by the cosine similarity of their embeddings. OpenCLIP uses the GPT-2 architecture~\cite{gpt2} for the text encoder and offers a broad spectrum of image encoders with convolutional (ConvNeXt~\cite{convnext}) and vision transformer (ViT~\cite{vit}) architectures. BLIP uses the BERT~\cite{BERT} text encoder and offers a base (BLIP-B) and a large (BLIP-L) large vision transformer as image encoders.
To ensure fair comparisons, we only create cascades of models with strictly increasing Recall@k \emph{and} computational costs.

\vspace{-10pt}
\paragraph{Results}
\Cref{tab:results} shows our results. 
For Flickr30k, the uncascaded ConvNeXt models B and L have up to 9.9x lower lifetime costs than the larger ConvNeXt-XXL. However, this comes at a price of significantly lower search quality --- R@k drops by up to 5.8\%. In contrast, the 2-level cascade [L, XXL] lowers lifetime costs by 3.1x at no loss in search quality. [B, XXL] improves this to 5.0x.

As noted in \cref{sec:cascadedsearch}, cascades with more than two levels offer no reduced lifetime costs but may reduce early query latency. To demonstrate this, we ``sandwich'' the medium-cost ConvNeXT-L between the cheap  ConvNeXT-B and the expensive ConvNeXT-XXL to obtain the 3-level cascade [B, L, XXL]. \Cref{alg:cascadedsearch} requires parameters $m_1$ and $m_2$ for a 3-level cascade. We keep $m_1 = 50$ to fairly compare against the 2-level cascade [B, XXL]. $m_2$ is a knob that yields larger savings for smaller values at a potential loss in search quality. We use \Cref{eq:latencyreduction} to compute $m_2 = 14$ as the value for which $F_{\textrm{latency}} \approx 2$ and use this $m_2$ for all 3-level cascades. \Cref{tab:results} shows that [B, L, XXL] indeed lowers early query latency by a factor of 1.97x and no significant change in search quality at slightly higher lifetime costs.

Results for ViT models and MSCOCO are similar and match the above analysis.

% \label{sec:2lvlcascades}

% \begin{table}[]
% \begin{tabular}{lllll}
% Model & R@1 & R@5 & R@10 & Cost \\ \hline
% ViT-B/32 & 65.2 & 87.5 & 92.9 & 9.4 GMAC \\
% ViT-B/16 & 69.1 & 88.8 & 94.2 & 3.8x \\
% ViT-L/14 & 71.6 & 91.1 & 95.0 & 18x \\
% ViT-g-14 & 75.6 & 92.6 & 95.8 & 61x \\ \hline\hline
% ConvNeXt-B &  &  &  & 0.6 TMAC \\
% ConvNeXt-L & 73.8 & 92.2 & 96.1 & 2.2x \\
% ConvNeXt-XXL & 75.0 & 93.8 & 96.4 & 10x \\ \hline\hline
% BLIP-B & 85.7 & 97.2 & 98.9 & 43 TMAC \\
% BLIP-L & 83.3 & 96.3 & 98.0 & 3.5x
% \end{tabular}
% \caption{}
% \label{tab:my-table}
% \end{table}

% Please add the following required packages to your document preamble:
% \usepackage[table,xcdraw]{xcolor}
% If you use beamer only pass "xcolor=table" option, i.e. \documentclass[xcolor=table]{beamer}
\begin{table}[]
\begin{tabular}{lllll}
\multicolumn{5}{@{}c}{Flickr30k}\\
\toprule
Cascade                & R@1  & R@5  & R@10 & \begin{tabular}[c]{@{}l@{}}$F_{\textrm{life}}$ ($F_{\textrm{latency}}$)\end{tabular}\\ \midrule \\[-7pt] 
\multicolumn{5}{c}{CLIP, ViT}                                                                                              \\[3pt]
\rowcolor[HTML]{EFEFEF} 
{[}g/14{]}             & 75.6 & 92.6 & 95.8 & 1x                                                                           \\
{[}B/16{]}             & -6.5 & -3.8 & -1.6 & 15.8x                                                                        \\
\rowcolor[HTML]{EFEFEF} 
{[}L/14{]}             & -4.0 & -1.7 & -0.8 & 3.4x                                                                         \\
{[}L/14, g/14{]}       & 0.0  & 0.0  & 0.0  & 2.6x                                                                         \\
\rowcolor[HTML]{EFEFEF} 
{[}B/16, g/14{]}       & 0.0  & -0.1 & -0.2 & 6.1x                                                                         \\
{[}B/16, L/14,\\\,\,g/14{]} & 0.0  & -0.4 & +0.1 & 5.2x (1.75x)                                                                      \\ \hline\\[-7pt]
\multicolumn{5}{c}{CLIP, ConvNeXt}                                                                                         \\[3pt]
\rowcolor[HTML]{EFEFEF} 
{[}XXL{]}              & 75.0 & 93.8 & 96.4 & 1x                                                                           \\
{[}B{]}                & -5.8 & -4.8 & -2.5 & 9.9x                                                                         \\
\rowcolor[HTML]{EFEFEF} 
{[}L{]}                & -1.2 & -1.6 & -0.3 & 4.4x                                                                         \\
{[}L, XXL{]}           & 0.0  & -0.4 & +0.4 & 3.1x                                                                         \\
\rowcolor[HTML]{EFEFEF} 
{[}B, XXL{]}           & +1.4 & -0.1 & -0.1 & 5.0x                                                                         \\
{[}B,  L, XXL{]}       & +0.4 & -0.3 & +0.3 & 4.5x (1.97x) \\
\bottomrule
\end{tabular}
\bigskip

\begin{tabular}{lllll}
\multicolumn{5}{@{}c}{MSCOCO}\\
\toprule
Cascade & R@1 & R@5 & R@10 & \begin{tabular}[c]{@{}l@{}}$F_{\textrm{life}}$ ($F_{\textrm{latency}}$)\end{tabular} \\ \midrule\\[-7pt]
\multicolumn{5}{c}{CLIP, ViT} \\[3pt]
\rowcolor[HTML]{EFEFEF} 
{[}g/14{]} & 46.3 & 70.68 & 79.84 & 1x \\
{[}B/16{]} & -6.2 & -5.3 & -4.5 & 15.8x \\
\rowcolor[HTML]{EFEFEF} 
{[}L/14{]} & -2.9 & -2.6 & -1.9 & 3.4x \\
{[}L/14, g/14{]} & +0.0 & +0.0 & +0.1 & 2.6x \\
\rowcolor[HTML]{EFEFEF} 
{[}B/16, g/14{]} & +0.1 & +0.1 & +0.1 & 6.1x \\
\begin{tabular}[c]{@{}l@{}}{[}B/16, L/14,\\\,g/14{]}\end{tabular} & +0.1 & +0.6 & +0.2 & 5.2x (1.75x) \\ \hline\\[-7pt]
\multicolumn{5}{c}{CLIP, ConvNeXt} \\[3pt]
\rowcolor[HTML]{EFEFEF} 
{[}XXL{]} & 45.78 & 70.74 & 79.60 & 1x \\
{[}B{]} & -7.3 & -6.3 & -5.1 & 9.9x \\
\rowcolor[HTML]{EFEFEF} 
{[}L{]} & -2.2 & -2.18 & -1.7 & 4.4x \\
\rowcolor[HTML]{EFEFEF} 
{[}B, XXL{]} & -0.3 & +0.3 & -0.2 & 5.0x \\
{[}B,  L, XXL{]} & +0.4 & -0.1 & -0.6 & 4.5x (1.97x) \\ \hline\\[-7pt]
\multicolumn{5}{c}{BLIP} \\[3pt]
\rowcolor[HTML]{EFEFEF} 
{[}L{]} & 60.05 & 83.78 & 90.56 & 1x \\
{[}B{]} & -0.5 & +0.1 & -0.3 & 3.5x \\
\rowcolor[HTML]{EFEFEF} 
{[}B, L{]} & 0.0 & +0.2 & +0.1 & 2.6x\\
\bottomrule
\end{tabular}
\\
\vspace{5pt}
\caption{Recall@$k$ in \% and cost reduction factors for various uncascaded models, 2-level cascades and 3-level cascades on Flickr30k and MSCOCO with $m_1 = 50$ and $m_2 = 14$. 2-level cascades list the lifetime cost reduction factor $F_{\textrm{life}}$. 3-level cascades list the early query latency cost reduction factor $F_{\textrm{latency}}$ relative to the preceding 2-level cascade. We omit results for BLIP on Flickr30k because BLIP-L does not improve over BLIP-B.}
\label{tab:results}
\end{table}

\section{Conclusion}
% \vspace{-5pt}
We have introduced model cascades that  reduce the lifetime computational search costs of BE TIR systems under the $p$-small-world search scenario. 
Our experiments show that \Cref{alg:cascadedsearch} can lower costs by over 6x at no reduction in search quality. At the same time,  deeper model cascades can mitigate the increase in latency of early queries, which is especially important for expensive modern neural encoders.
Anecdotal evidence supports our choice of $p = 10\%$. With that said, practical search scenarios likely vary in $p$ and achieve accordingly different speedups.

Single-digit speedups may not sufficiently reduce computational costs to economically rank large-scale image databases with expensive neural BEs. Instead, a practitioner may use traditional search engines to retrieve the top-$k$ images and apply a neural search cascade on top of it. This heterogeneous cascade may offer a viable path towards integrating state-of-the-art neural networks with established image search platforms.

{\small
\bibliographystyle{ieee_fullname}
\bibliography{egbib}
}

\end{document}

%% file: mathcommands.tex
\def\eqref#1{equation~\ref{#1}}
\def\1{\bm{1}}

\def\vv{{\bm{v}}}

\def\mV{{\bm{V}}}

\DeclareMathAlphabet{\mathsfit}{\encodingdefault}{\sfdefault}{m}{sl}
\SetMathAlphabet{\mathsfit}{bold}{\encodingdefault}{\sfdefault}{bx}{n}

\def\gD{{\mathcal{D}}}

\def\gI{{\mathcal{I}}}

\def\sN{{\mathbb{N}}}

\newcommand{\bracket}[1]{\left[#1\right]}
\renewcommand{\brace}[1]{\left(#1\right)}
\newcommand{\cbrace}[1]{\left\{#1\right\}}

\newcommand{\abs}[1]{\left|#1\right|}

\newcommand{\f}[2]{#1\!\brace{#2}}

%% file: images/architecture.tikz
\begin{tikzpicture}[
image/.style={rectangle, draw=none},
small/.style={rectangle, draw=none, fill=blue!20, rounded corners=2mm},
large/.style={rectangle, draw=none, fill=blue!40, rounded corners=2mm},
image_encoder/.style={minimum height=10mm, minimum width=10mm},
text_encoder/.style={minimum height=10mm, minimum width=10mm},
box/.style={rectangle, draw=white, fill=red!70, minimum width=2mm, minimum height=2mm},
CLIP/.style={dashed, draw=violet },
BUILD/.style={dotted, draw=black,thick,rounded corners=2mm},
RUNTIME/.style={dotted, draw=black,thick,rounded  corners=2mm},
arrow/.style={->,thick,blue!70},
bar/.style={thick, blue!70},
]

% \draw[help lines,step=5mm,gray!20] (0,0) grid (12,8);
\foreach \src/\x in {000000000285.jpg/0.0, 000000001584.jpg/0.1, 000000001584.jpg/0.2, 000000006040.jpg/0.3}{
    \node[image] () at (\x, \x) {\includegraphics[width=20mm, height=10mm]{images/\src}};
}

\node[small, image_encoder] at (30mm, 1mm) (small_image_encoder)  {$I_s$};
% \node[small, text_encoder] at (45mm, 20mm) (small_text_encoder)  {$\tilde E^{\mathcal{T}}$};

\node[large, image_encoder] at (60mm, 1mm) (small_image_encoder)  {$I_l$};
\node[large, text_encoder] at (50mm, 15mm) (small_text_encoder)  {$T$};

\foreach \y in {-0.2,0.0,...,0.4}{
% image to the first CLIP
    \node[box] at (45mm, \y){};
    \draw [arrow] (13mm, \y) -- (25mm,\y);
    \draw [arrow] (35mm, \y) -- (44mm,\y);
}
\draw [arrow] (50mm, 10mm) --  (45mm,5mm);
\foreach \y in {0.2,0.4}{
%first CLIP to the second CLIP
    \node[box, fill=green] at (45mm, \y){};
    \node[box, fill=green] at (50mm, \y){};
    \draw [arrow] (46mm, \y) -- (49mm,\y);
    \draw [arrow] (51mm, \y) -- (55mm,\y);
    \node[box] at (75mm, \y){};
    \draw [arrow] (65mm, \y) -- (74mm,\y);
}
\draw [arrow] (50mm, 10mm) -- (75mm,5mm);
\node[box, fill=green] at (75mm, 4mm){};
\node[box, fill=green] at (80mm, 4mm){};
\draw [arrow] (76mm, 4mm) -- (79mm,4mm);
\draw [arrow] (81mm, 4mm) -- (85mm,4mm);

\node[] at (50mm, 25mm) {``A photo of a tram."};
\foreach \x in {5}{
        \draw[arrow] (50mm, 23mm) -- (\x, 20mm);
}

\draw[arrow] (0mm, -5mm) -- (0mm, -10mm) -- (50mm, -10mm) -- (50mm, 1mm);
\draw[arrow] (50mm, -10mm) -- (80mm, -10mm) -- (80mm, 3mm);

% \draw[CLIP] (23mm, -15mm) rectangle (52mm, 30mm);
% \draw[CLIP] (53mm, -15mm) rectangle (82mm, 30mm);
% \node[text=violet] at (45mm,-10mm) {$f_1$};
% \node[text=violet] at (75mm,-10mm) {$f_2$};
\draw[BUILD] (20mm, -15mm) rectangle (37mm, 20mm);
\draw[RUNTIME] (38mm, -15mm) rectangle (85mm, 20mm);
\node[text=black] at (28mm, -13mm) {build time};
\node[text=black!100] at (60mm, -13mm) {run time};

\end{tikzpicture}